\documentclass[aps,prd,twocolumn,groupedaddress,showpacs,nofootinbib]{revtex4}
\usepackage{graphicx}

\newcommand{\be}{\begin{equation}}
\newcommand{\ee}{\end{equation}}
\newcommand{\bea}{\begin{eqnarray}}
\newcommand{\eea}{\end{eqnarray}}
\newcommand{\beaa}{\begin{eqnarray*}}
\newcommand{\eeaa}{\end{eqnarray*}}

\newcommand{\nn}{\nonumber \\}
\newcommand{\e}{{\rm e}}

\begin{document}

\title{Observational constraints on dark energy with generalized equations of state}

\author{S. Capozziello$^1$\thanks{Electronic address: capozziello@na.infn.it}, V.F. Cardone$^2$\thanks{winny@na.infn.it}, E. Elizalde$^3$\thanks{Electronic address: elizalde@ieec.uab.es}, S. Nojiri$^4$\thanks{Electronic address: nojiri@cc.nda.ac.jp}, S.D. Odintsov$^3$\thanks{Electronic address: odintsov@ieec.uab.es}}

\affiliation{$^1$Dipartimento di Scienze Fisiche, Universit\`{a} di Napoli ``Federico II'' and INFN, Sez. di Napoli, Compl. Univ. Monte S. Angelo, Edificio N, Via Cinthia, I-80126, Napoli, Italy,\\
$^2$Dipartimento di Fisica "E.R. Caianiello", Universit\`{a} di Salerno, and INFN, Sez. di Napoli, Via S. Allende, I-84081, Baronissi (Salerno), Italy,\\
$^3$Instituci\`{o} Catalana de Recerca i Estudis Avan\c{c}ats (ICREA) and Institut d' Estudis Espacials de Catalunya (IEEC/ICE),
Edifici Nexus, Gran Capit\`{a} 2-4, 08034 Barcelona, Spain,\\
$^4$Department of Applied Physics, National Defence Academy, Hashirimizu Yokosuka 239-8686, Japan}

\begin{abstract}

We investigate  the effects of viscosity terms depending on the Hubble parameter and its derivatives in the dark energy equation of state. Such terms are possible if  dark energy is a fictitious fluid originating from corrections to the Einstein general relativity as is the case for some braneworld inspired models or fourth order gravity. We consider two classes of models whose singularities in the early and late time universe have been studied by testing the models against the dimensionless coordinate distance to Type Ia Supernovae and radio-galaxies also including priors on the shift and the acoustic peak parameters. It turns out that both models are able to explain the observed cosmic speed up without the need of phantom $(w < -1)$ dark energy.

\end{abstract}

\pacs{98.80.-k, 98.80.Es, 97.60.Bw, 98.70.Vc}

\maketitle

\section{Introduction}

According to the new cosmological picture emerging from data (only few years ago unexpected), we live in a spatially flat universe filled with a subcritical matter content and undergoing an accelerated expansion. The Hubble diagram of Type Ia Supernovae (hereafter SNeIa) \cite{SNeIa} has been the first cornerstone, but quite soon other observational data, from the cosmic microwave background (hereafter CMBR) anisotropy spectrum \cite{CMBR} to the large scale structure properties \cite{LSS}, further corroborate the first impression from SNeIa. The astonishingly precision of the CMBR spectrum measured by the WMAP satellite \cite{WMAP} and the Gold SNeIa sample of Riess et al. \cite{Riess04} represent the last and still more convincing evidences in favor of this new description of the universe.

Both the cosmic speed up and the flatness of the universe have posed serious problems to the cosmologist community. Since matter alone (both visible and dark) cannot be enough to close the universe, a new component was invoked as a dominant term. Moreover, in order to explain the observed cosmic speed up, this new fluid must have a negative pressure. Being obscure both in its origin and its properties, this component was baptized {\it dark energy}. Understanding its nature and nurture is one of the most fascinating and debated challenges of modern cosmology. Although the old cosmological constant \cite{Lambda} may play naturally the role of dark energy and also fits well the observed dataset \cite{Teg03,Sel04}, it is affected by serious theoretical shortcomings that have motivated the search for alternative candidates. It is nevertheless worth observing that, to a large extent, the different models may be broadly classified in three classes which we will refer to as {\it scalar fields}, {\it unified dark energy} (hereafter UDE) and {\it modified gravity}. (Sometimes, some of these models maybe rewritten as models from another class).

For models belonging to the first class, dark energy is an effective fluid originating from an ultralight scalar field $\phi$, dubbed {\it quintessence}, evolving under the action of a suitably chosen self\,-\,interaction potential $V(\phi)$. The choice of $V(\phi)$ is crucial and different functional expressions have been investigated, from power\,-\,law \cite{RP} to exponentials \cite{exp,RS} and a combination of both \cite{SUGRA}. Without entering into details (for which the reader is referred to \cite{QuintRev}), we only stress that such models are able to correctly reproduce the observed data, but are affected by fine tuning problems and the open issue of understanding where the quintessence scalar field comes from.

Notwithstanding the strong efforts made up to now, it is not known what is the fundamental nature of the dark energy so that it is also worth investigating the possibility that dark energy and dark matter (the other unknown component of the universe) are two different aspects of the same  fluid. This is the underlying idea of models belonging to the UDE class. In such an approach, a single fluid with an exotic equation of state plays the role of dark matter at high densities and dark energy at low densities. Typical examples are the Chaplygin gas \cite{Chaplygin}, the tachyonic fluid \cite{tachyon} and the Hobbit \cite{Hobbit} model\footnote{Similar to the UDE models is the phenomenological {\it inflessence} scenario \cite{ie} where a single fluid is used to explain both inflation and dark energy.}.

In both classes described above, dark energy is explained in terms of a new fluid in the framework of standard general relativity. However, the Einstein theory of gravity has been experimentally tested only on scales up to the Solar System and hence it is far from be verified that its validity holds also on cosmological scales. Motivated by these considerations, different models have been proposed where the cosmic speed up is explained in terms of a matter only universe regulated by dynamical equations that are different from the usual Friedmann ones as a consequence of a generalized gravity theory. Examples are the braneworld inspired Dvali\,-\,Gabadadze\,-\,Porrati model \cite{DGP} and fourth order theories of gravity, both in the metric \cite{capozcurv,review,noicurv,MetricRn} and Palatini \cite{PalRn,lnR,Allemandi} formulation. Although such models are able to give rise to accelerated expansion without the need of any dark energy, a strong debate is still open about their compatibility with the standard tests of gravity in the low energy limit.

From the short (and far from exhaustive) overview presented above, it is clear that there is much confusion. However, it is possible to put some order in this somewhat chaotic situation by considering a particular feature of the dark energy, namely its equation of state (hereafter EoS), i.e. the relation between the pressure and the energy density. Indeed, whatever is the model considered, it is always possible to introduce a sort of dark energy fluid whose energy density and pressure are determined by the characteristics of the given theory. While for scalar fields and UDE models, the EoS is a function of the energy density only, in the case of modified gravity theories,  the effective EoS depends also on geometry, e.g. on the Hubble parameter and/or its derivatives. It is therefore tempting to investigate the properties of cosmological models starting from the EoS directly and testing whether a given EoS is able to give rise to cosmological models reproducing the available dataset. A first step forward in this direction has been undertaken  in a recent paper \cite{NO05} where the singularities of models assigned by generalized EoS have been investigated. Since the main interest was there in asymptotic behaviors (i.e., in the far past and far future), any matter term was not included in the analysis in \cite{NO05} which makes it possible to analytically solve the dynamical equations. In order to elaborate further on this original idea, we have to include the contribution of dust (dark/baryonic)  matter. The present work is thus complementary to \cite{NO05} since $i.)$ we consider the present rather than the asymptotic states of the models, and $ii.)$ we investigate the viability of the models and constrain their parameters using the observational data.

The plan of the paper is as follows. In Sect.\,II, we introduce the models that we investigate, chosen for their interesting asymptotic behaviors and give a general theoretical discussion on generalized EoS. The observational data, the method used to constrain the model parameters and the results of the analysis are presented in Sect.\,III. Some hints on the high redshift behaviour of the models are presented in Sect.\,IV, while, in Sect.\,V, we qualitatively discuss the issue of structure formation. We summarize and conclude in Sect.\,VI, while further details on the theoretical foundation of the generalized EoS we consider are given in two appendixes.

\section{The models}

For a spatially flat homogenous and isotropic Robertson\,-\,Walker universe, the Einstein equations reduce to the usual Friedmann equations\,:

\begin{equation}
H^2 = \frac{\kappa^2}{3} \rho \ ,
\label{eq: fried1}
\end{equation}

\begin{equation}
2 \dot{H} = - \kappa^2 (\rho + p)
\label{eq: fried2}
\end{equation}
where $H = \dot{a}/{a}$ is the Hubble parameter, $a$ the scale factor and an overdot denotes the derivative with respect to cosmic time $t$. On the right hand side of Eqs.(\ref{eq: fried1}) and (\ref{eq: fried2}),  we have set $\kappa^2 = 8 \pi G$, while $\rho$ and $p$ are the total energy density and pressure respectively. Since the radiation term is nowadays negligible, we assume that the universe is filled only with dust matter and a dark energy  fluid and use subscripts $M$ and $X$ to denote quantities referring to the first and second component respectively. Provided the EoS $p_i = p_i(\rho_i)$ is somewhat given, the evolution of the energy density of the $i$\,-\,th fluid may be (at least, in principle) determined by solving the continuity equation\,:

\begin{equation}
\dot{\rho}_i + 3 H (\rho_i + p_i) = 0 \ .
\label{eq: cont}
\end{equation}
This may be conveniently rewritten in terms of the redshift $z = 1/a - 1$ (having assumed $a = 1$ at the present day) as\,:

\begin{equation}
\frac{d\rho_i}{dz} = \frac{3 (1 + w_i) \rho_i}{1 + z}
\label{eq: contz}
\end{equation}
with $w_i \equiv p_i/\rho_i$ the EoS parameter\footnote{In the following, as {\it EoS} we will refer both to the relation $p_i = p_i(\rho_i)$ and to $w_i$ indifferently.}. For dust matter, $w_M = 0$, Eq.(\ref{eq: contz}) is easily integrated to give $\rho_M = \Omega_M \rho_{crit} (1 + z)^3$ with $\Omega_M = \rho_M/\rho_{crit}$ the matter density parameter, $\rho_{crit} = 3 H_0^2/\kappa^2$, the present day critical density and hereon quantities with a subscript $0$ are evaluated at $z = 0$.

Concerning the dark energy, almost nothing is known about its EoS given the ignorance of its nature and its fundamental properties. The simplest choice

\begin{equation}
p_X = w_X \rho_X \,,\label{eq: constw}
\end{equation}
with $w_X$ a constant, has the virtue of leading to an integrable continuity equation thus yelding\,:

\begin{equation}
\rho_X = \Omega_X \rho_{crit} (1 + z)^{3 (1 + w_X)}
\label{eq: constwrho}
\end{equation}
with $\Omega_X$ the dark energy density parameter. Note that, because of the flatness condition, it is $\Omega_M + \Omega_X = 1$.

Although a constant EoS dark energy model (also referred to as {\it quiessence} or QCDM) nicely fits the available data, there are two serious shortcomings in this approach. First, there are no theoretical models predicting a rigorously constant EoS so that this assumption lacks a whatever background motivation. Moreover, fitting a large set of data points towards $w_X < -1$ as best fit (see, e.g., \cite{EstW}) so that $\rho_X + p_X < 0$. Models having $w_X < -1$ are collectively referred to as {\it phantom} models \cite{phantom} and have some disturbing features such as violating the weak energy condition $\rho_X + p_X \ge 0$ and leading to a divergence of the scale factor $a(t)$ in a finite time (referred to as {\it Big Rip}).

It is worth noting that allowing $w_X$ to evolve with $z$ does not seem to solve the problem. On the contrary, fitting parametrized model independent EoS to the available dataset, as $w_X = w_0 + w_1 z$ \cite{linz} or $w_X = w_0 + w_a z/(1 + z)^p$ with $p = 1, 2$ \cite{lina}, still points at $w_0 < -1$. Moreover, the results seem to suggest that $w_X$ crosses the {\it phantom divide}, i.e. the EoS changes from $w_X > -1$ in the past to $w_X < -1$ today. Recently, this feature has been recovered in exact models where inflation is matched to late-time acceleration by scalar phantom\,-\,non\,-\,phantom transitions \cite{uni}. Other realistic models are achieved by introducing also dark matter into the game \cite{any}.

Here, we investigate the possibility that the dark energy EoS depends not only on the energy density $\rho_X$, but also on the Hubble parameter $H$ and/or its derivatives. We will refer to such a model as {\it generalized EoS}. As a preliminary remark, let us note that these same models were also referred to as {\it inhomogeneous EoS} in \cite{NO05}. However, it is worth stressing that the term {\it inhomogeneous} does not mean that we consider an inhomogeneous universe (so that the RW metric may still be retained) nor that we consider the possibility that dark energy may cluster in the nonlinear regime of perturbations (as, e.g., in \cite{clusde}). We use the term {\it inhomogeneous} since assuming $w_X = w_X(\rho_X, H, \dot{H})$ introduces a viscosity term in the continuity equation similar to what happens in fluidodynamics for an inhomogeneous fluid. See Appendix A for an explicit derivation, while specific examples of EoS due to time\,-\,dependent bulk viscosity are given in, e.g., \cite{brevik}.

In the following, we will describe two  classes of models introduced in \cite{NO05} chosen because of their interesting asymptotic properties. Actually, we slightly generalize them and find out some degeneracies from the point of view of observations. Finally, we discuss the general theoretical features that a given generalized EoS should fulfill in order to match with any observational cosmology.

\subsection{Increased Matter}

Let us first consider the case\,:

\begin{equation}
p_X = w_f \rho_X + w_H H^2 \label{eq: wa}
\end{equation}
with $(w_f, w_H)$ two undetermined constants\footnote{Note that, while $w_f$ is dimensionless, the dimensions of $w_H$ are fixed in such a way that $w_H H^2$ is a pressure term.}. Eq.(\ref{eq: wa}) is a particular case of the more general class of dark energy models with EoS\,:

\begin{equation}
p_X = - \rho_X - A \rho_X^{\alpha} - B H^{2 \beta} \label{eq:
wagen}
\end{equation}
that reduces to Eq.(\ref{eq: wa}) by setting $\alpha = \beta = 1$, $w_f = -(1 + A)$ and $B = w_H$. The singularities (in the far past and in the far future when the matter term may be neglected) of this class of models have been investigated in detail in \cite{NO05} where it has been shown that a wide range of interesting possibilities may be achieved depending on the values of $(\alpha, \beta)$. In particular, for $\alpha = \beta = 1$, even if $w_0 < -1$, the Big Rip is avoided provided that $w_{eff} = w_f + w_H \kappa^2/3 > -1$ (for general classification of future DE singularities, see\cite{tsujikawa}). While the general model in Eq.(\ref{eq: wagen}) is characterized by five parameters (namely, $\alpha$, $\beta$, $A$, $B$ and $\Omega_M$) and is therefore quite difficult to constrain observationally, the case in Eq.(\ref{eq: wa}) has the virtue of avoiding the Big Rip and (possibly) the need for phantom fields with the further advantage of being assigned by only three parameters. For these reasons, we will consider hereon only this particular realization of Eq.(\ref{eq: wagen}).

The EoS (\ref{eq: wa}) may be written in a more significant way using Eq.(\ref{eq: fried1}) to express $H^2$ as function of $\rho_X$ and $\rho_M$. We thus get\,:

\begin{eqnarray}
p_X & = & w_f \rho_X + w_H H_0^2 \left [ \Omega_M (1 + z)^3 + \eta(z) \right ] \nonumber \\
~ & ~ & ~ \nonumber \\
~ & = & \displaystyle{\left [ w_{DE} + \frac{w_{M}^{eff} \Omega_M (1 + z)^3}{\eta(z)} \right ] \rho_X(z)} \ ,
\label{eq: pa}
\end{eqnarray}
so that the dark energy EoS reads\,:

\begin{equation}
w_X =  w_{DE} + \frac{w_{M}^{eff} \Omega_M (1 + z)^3}{\eta(z)} \ .
\label{eq: wavsz}
\end{equation}
In Eq.(\ref{eq: pa}), in the first row, we have rewritten Eq.(\ref{eq: fried1}) as $H^2/H_0^2 = E^2 = \Omega_M (1 + z)^3 + \eta(z)$ with $\eta \equiv \rho_X/\rho_{crit}$, while, in the second row, we have defined\,:

\begin{equation}
\left \{
\begin{array}{l}
w_{DE} = \displaystyle{w_f + \frac{\kappa^2 w_H}{3}} \nonumber \\
~ \\
w_{M}^{eff} = \displaystyle{\frac{\kappa^2 w_H}{3}} \nonumber \\
\end{array}
\right . \ .
\label{eq: defwde}
\end{equation}
Eq.(\ref{eq: wavsz}) nicely shows that the model we are considering is a simple generalization of the QCDM case to which it reduces when $w_H = 0$. For $w_H \ne 0$, $w_X(z)$ may still be expressed analytically. To this aim, let us insert Eq.(\ref{eq: wavsz}) in the continuity equation for $\rho_X$ and integrate it with the initial condition $\rho_X(z = 0) = \Omega_X \rho_{crit}$. Using the dimensionless quantity $\eta = \rho_X/\rho_{crit}$, we find\,:

\begin{equation}
\eta = \left ( \Omega_X + \frac{w_M^{eff} \Omega_M}{w_{DE}} \right ) (1 + z)^{3 (1 + w_{DE})}
 - \frac{w_M^{eff} \Omega_M}{w_{DE}} (1 + z)^3 \ .
\label{eq: etamoda}
\end{equation}
Inserting now Eq.(\ref{eq: etamoda}) into Eq.(\ref{eq: fried1}), we obtain\,:

\begin{equation}
E^2(z) = \tilde{\Omega}_M (1 + z)^3 + (1 - \tilde{\Omega}_M) (1 + z)^{3 (1 + w_{DE})}
\label{eq: ea}
\end{equation}
with

\begin{equation}
\tilde{\Omega}_M = \Omega_M \left (1 - \frac{w_M^{eff}}{w_{DE}} \right ) \ .
\label{eq: deftom}
\end{equation}
We thus get a quite interesting result. The EoS (\ref{eq: wavsz}) leads to the Hubble parameter that has formally the same expression as that of QCDM models, but at the price of shifting both the matter density parameter and the dark energy EoS. As such, a fitting analysis based on observables depending only on $E(z)$ as is the case of the SNeIa Hubble diagram (see later) will give biased results. In particular, if $w_{DE} < 0$ and $w_H > 0$, then $\tilde{\Omega}_M > \Omega_M$ so that the best fit values $(w_{DE}^{fit}, \Omega_M^{fit})$ may erroneously point towards a universe with a higher matter content and a  phantom dark energy. Because of this peculiarity, we will refer in the following to this scenario as the {\it increased matter} (hereafter $IM$) model, even if the increasing of the matter content is only apparent.

It is interesting to note that, fitting only the SNeIa Gold dataset without any prior on $\Omega_M$, Jassal et al. \cite{JBP05} have found $\Omega_M = 0.47$ as best fit, significantly higher than the fiducial value $\Omega_M \simeq 0.3$ suggested by cluster abundances. On the other hand, fitting the WMAP data only gives $\Omega_M = 0.32$ as best fit value thus rising the problem of reconciling these two different estimates. In their conclusions, Jassal et al. argue in favor of possible systematic errors in one of the dataset. Eqs.(\ref{eq: ea}) and (\ref{eq: deftom}), however, furnish an alternative explanation. Simply, since (as we will see more in detail later) the SNeIa Hubble diagram only probes $E(z)$, fitting this dataset with no priors gives constraints on $\tilde{\Omega}_M$ rather than $\Omega_M$. Since $\tilde{\Omega}_M > \Omega_M$, Eq.(\ref{eq: deftom}) suggests that a a possible way to reconcile this worrisome discrepancy may be to resort to our model assuming $w_{M}^{eff}/w_{DE} < 0$. Since it is reasonable to expect that $w_{DE} < 0$, $w_H > 0$ have to imposed.

Provided a suitable method is used to determine $\Omega_M$, $w_{DE}$ and $\tilde{\Omega}_M$, it is possible to get an estimate of $w_M^{eff}$ as

\begin{equation}
w_M^{eff} = \frac{w_{DE} (\Omega_M - \tilde{\Omega}_M)}{\Omega_M}
\label{eq: estwm}
\end{equation}
and then of the {\it true} barotropic factor $w_f$ in Eq.(\ref{eq: wavsz}) as\,:

\begin{equation}
w_f = w_{DE} - w_{M}^{eff}
\label{eq: estwf}
\end{equation}
which shows that, assuming $w_H > 0$ for what said above, leads to $w_f < w_{DE}$. As a consequence, if the effective barotropic factor $w_{DE}$ is in the phantom region, the true one $w_f$ will be deeper in the phantom. As such, the model is thus not able to evade the need for phantom dark energy, but it is still interesting since it makes it possible to avoid any Big Rip if $w_{DE} + w_M^{eff}$ is larger than the critical value ($w_X = -1$).

There is, actually, an easy generalization of the model able to both eliminate future singularities and phantom fields. To this aim, let us consider the following EoS\,:

\begin{equation}
p_X = w_f \rho_X + w_H H^2 + w_{dH} \dot{H} \label{eq: wabis}
\label{eq: padot}
\end{equation}
with $w_{dH}$ a new constant. Eq.(\ref{eq: wabis}) is inspired by Eq.(47) in \cite{NO05} to which it reduces by setting $w_f = w$, $w_H = -3(1 + w)/\kappa^2$ and $w_{dH} = -2/\kappa^2$. It is easy to show that, with these positions and in absence of matter, Eq.(\ref{eq: wabis}) is an identity. When a matter term is added and the constants $(w_f, w_H, w_{dH})$ are left free, this is no more true and interesting consequences come out. To see this, let us use the Friedmann equations to get $H^2$ and $\dot{H}$ in terms of $(\rho_M, \rho_X, p_X)$. It is then easy to show that Eq.(\ref{eq: wabis}) may be rewritten as\,:

\begin{eqnarray}
\left ( 1 + \frac{w_{dH} \kappa^2}{2} \right ) p_X & = & \left ( w_f + \frac{w_H \kappa^2}{3} - \frac{w_{dH} \kappa^2}{2} \right ) \rho_X
\nonumber \\
~ & + & \left ( \frac{w_H \kappa^2}{3} - \frac{w_{dH} \kappa^2}{2} \right ) \rho_M
\label{eq: pabis}
\end{eqnarray}
so that the dark energy EoS is given again by Eq.(\ref{eq: wavsz}) provided Eqs.(\ref{eq: defwde}) are generalized as\,:

\begin{equation}
\left \{
\begin{array}{l}
w_{DE} = \displaystyle{\left ( w_f + \frac{\kappa^2 w_H}{3} - \frac{w_{dH} \kappa^2}{2} \right )
\left ( 1 + \frac{w_{dH} \kappa^2}{2} \right )^{-1}} \nonumber \\
~ \\
w_{M}^{eff} = \displaystyle{\left ( \frac{\kappa^2 w_H}{3} - \frac{w_{dH} \kappa^2}{2} \right )
\left ( 1 + \frac{w_{dH} \kappa^2}{2} \right )^{-1}} \nonumber \\
\end{array}
\right .
\label{eq: defwdegen}
\end{equation}
while $\tilde{\Omega}_M$ is still defined by Eq.(\ref{eq: deftom}) with $w_{DE}$ and $w_{M}^{eff}$ now given by Eqs.(\ref{eq: defwdegen}). By using this generalization, we can still get $\tilde{\Omega}_M > \Omega_M$, but now the conditions $w_{M}^{eff}/w_{DE} > 0$ and $w_{DE} < 0$ does not imply $w_H > 0$. As a further consequence, the true barotropic factor $w_f$ is now given as\,:

\begin{equation}
 w_f = \left ( 1 + \frac{w_{dH} \kappa^2}{2} \right ) (w_{DE} - w_{M}^{eff})
\label{eq: estwfgen}
\end{equation}
and we can get $w_f > -1$ provided that the condition

\begin{equation}
w_{dH} > - \frac{2}{\kappa^2} \left ( \frac{w_{DE} - w_M^{eff} + 1}{w_{DE} - w_M^{eff}} \right ) \label{eq: condwdh}
\end{equation}
is verified. Since Eqs.(\ref{eq: pa}) and (\ref{eq: pabis}) lead to models that are fully equivalent from the dynamical point of view, we will consider hereon these models as a single one referred to as the {\it increased matter} model.

\subsection{Quadratic EoS}

Let us now consider a different approach to the dark energy EoS. As yet stated above, we may think at the EoS as an implicit relation such as\,:

\begin{displaymath}
 F(p_X, \rho_X, H) = 0
\end{displaymath}
which is not constrained to lead to a linear dependence of $p_X$ on $\rho_X$. As a particularly interesting example, let us consider the case\,:

\begin{equation}
(\rho_X + p_X)^2 - C_s \rho_X^2 \left (1 - \frac{H_s}{H} \right ) = 0
\label{eq: pb}
\end{equation}
with $C_s$ and $H_s$ two positive constants.  Eq.(\ref{eq: pb}) has been proposed in \cite{NO05} where it has been shown that the corresponding cosmological model presents both Big Bang and Big Rip singularities. As such, this model may be a good candidate to explain not only the present state of cosmic acceleration, but also the inflationary one and is therefore worth to be explored. Since its EoS is given by a quadratic equation in $w_X$, we will refer to it in the following as the {\it quadratic EoS} (hereafter $QE$) model.

As a first step, it is convenient to rewrite the continuity equation for $\rho_X$ as a first order differential equation for $E(z) = H(z)/H_0$. To this aim, one may differentiate both members of Eq.(\ref{eq: fried1}) with $\rho = \rho_X + \Omega_M \rho_{crit} (1 + z)^3$ and solve with respect to $d\rho_X/dz$. On the other hand, from Eq.(\ref{eq: pb}), we easily get\,:

\begin{equation}
(1 + w_X) \rho_X = {\pm} \rho_X \sqrt{C_s \left ( 1 - \frac{E_s}{E} \right )}
\label{eq: wb}
\end{equation}
where the sign changes for $w_X=-1$ (phantom divide), corresponding to the standard $\Lambda$CDM model. On the other hand, the sign changes also for $E_s=E$ which corresponds to a finite redshift $z=z_s$. The evolution of the Hubble parameter is achieved by inserting the previous expressions for $d\rho_X/dz$ and $(1 + w_X) \rho_X$ in Eq.(\ref{eq: contz}), we finally get\,:

\begin{eqnarray}
\frac{dE}{dz} & = & \displaystyle{\frac{3 \Omega_M (1 + z)^2}{2 E}} \nonumber \\
~ & \pm & \displaystyle{\frac{3\left [ E^2 - \Omega_M (1 + z)^3 \right ] \left [ C_s \left ( 1 -E_s/E \right ) \right ]^{1/2}}{2 E (1 + z)}}
\label{eq: deb}
\end{eqnarray}
which may be integrated numerically with the initial condition $E(z = 0) = 1$ provided that values of $(\Omega_M, E_s, C_s)$ are given. To this end, it is convenient to express $C_s$ in terms of a more common quantity. To this aim, let us remember that the deceleration parameter $q = - a \ddot{a}/\dot{a}^2$ is related to the dimensionless Hubble parameter $E(z)$ as\,:

\begin{equation}
q(z) = -1 + \frac{1 + z}{E(z)} \frac{dE(z)}{dz} \label{eq: qz}
\end{equation}
so that $(dE/dz)_{z = 0} = 1 + q_0$. Evaluating $(dE/dz)_{z = 0}$ from Eq.(\ref{eq: deb}) and solving with respect to $C_s$, we get\,:

\begin{equation}
C_s = \frac{\left [ 3 \Omega_M - 2 (1 + q_0) \right ]^2}{9 (1 - \Omega_M)^2 (1 - E_s)}
\label{eq: csqz}
\end{equation}
where no sign ambiguity is present. Hereon we will characterize the model by the values of the parameters $(q_0, E_s, \Omega_M)$. It is interesting to note that $C_s$ vanishes for $\Omega_M = 2(1 + q_0)/3$ which takes physically acceptable values for $-1 < q_0 < 1$. In such cases, Eq.(\ref{eq: pb}) reduces to $p_X = - \rho_X$, i.e. we recover the usual $\Lambda$CDM model for which it is indeed $q_0 = -1 + 3 \Omega_M/2$. On the other hand, $C_s$ seems to diverge for $\Omega_M = 1$. Actually, this is not the case. Indeed, for $\Omega_M = 1$, we have a matter only universe for which it is $q_0 = 1/2$ thus giving $C_s = 0$ and $E^2 = \Omega_M (1+ z)^3$ so that Eq.(\ref{eq: deb}) is identically satisfied. Once Eq.(\ref{eq: deb}) has been integrated, one may use Eq.(\ref{eq: fried1}) to get $\rho_X(z)$ and hence $w_X(z)$ from Eq.(\ref{eq: wb}).

Some important remarks are due at this point. If we start from a non\,-\,phantom phase $(1+w_X > 0)$ imposing $z=0$ as initial condition, for large $z$ (far past) the universe remains in a non\,-\,phantom phase. This is the case which we will match with observational data. On the other hand, starting from a phantom regime $(1+w_X < 0)$, we can reach a phase where $1+ w_X = 0$ or $E=E_s$ in a finite time, i.e. for finite $z$. In this case, $E(z)$ decreases with $z$, reaches $E_s$ for a finite $z_s$ and after starts to increase. The sign in Eq.(\ref{eq: wb}) hanges at $E=E_s$, then $E_s$ is a minimum for $E(z)$. This means that, for $z<z_s$, the sign of the second term of r.h.s of Eq.(\ref{eq: deb}) is minus, while, for $z>z_s$, the sign is plus. In conclusion, depending on the nature of dark energy fluid $(w_X)$, the model can represent a non\,-\,phantom cosmology or the evolution from a phantom to a non\,-\,phantom cosmology through the crossing of the phantom divide. For  sake of simplicity, we will match with observations only the non\,-\,phantom solution since we want to test the viability of generalized EoS but, in  principle, the method to constrain method which we will discuss below can be extended also to phantom cosmology.

\subsection{Generality of generalized EoS of the universe}

The above $IM$ and $QE$ models are particular examples of generalized EoS which can be used to fit observational data. A general approach to our generalized EoS is possible from a theoretical point of view. We will show that any observational cosmology may emerge from such EoS. By using a single function $f(t)$, we now assume the following EoS\,:

\be \label{SN1} p=-\rho -
\frac{2}{\kappa^2}f'\left(f^{-1}\left(\kappa\sqrt{\frac{\rho}{3}}\right)\right)\
. \ee
Then it is straightforward to show that a solution of the Friedmann equations (\ref{eq: fried1}), (\ref{eq: fried2}), and the continuity equation (\ref{eq: cont}) is given by

\be
\label{SN2} H=f(t)\ ,\quad \rho = \frac{3}{\kappa^2}f(t)^2\ ,\quad
p= \frac{3}{\kappa^2}f(t)^2 - \frac{2}{\kappa^2}f'(t)\ . \ee
Then {\it any} cosmology given by $H=f(t)$ can be realized by the EoS (\ref{SN1}).

As first example, if $f(t)$ and therefore $H$ is given by

\be
\label{SN3} f(t)=H=\frac{2}{3\left(w_m +
1\right)}\left(\frac{1}{t} + \frac{1}{t_s - t}\right) \ ,
\ee
the EoS has the following form\,:

\be \label{SN4} p=-\rho \mp \frac{(w_m
+ 1)\rho}{t_s} \sqrt{t_s^2 - \frac{8}{3(w_m +
1)\kappa}\sqrt{\frac{3}{\rho}}}\ .
\ee
Since

\be \label{any14}
\dot H=\frac{2}{3\left(w_m + 1\right)}\left(-\frac{1}{t^2} +
\frac{1}{\left(t_s - t\right)^2}\right)\ ,
\ee
the EoS parameter $w_{\rm eff}$ defined by

\be \label{FRW3k} w_{\rm
eff}=\frac{p}{\rho}= -1 - \frac{2\dot H}{3H^2}\ ,
\ee
goes to $w_m>-1$ when $t\to 0$ and goes to $-2 - w_m<-1$ at large times. The crossing $w_{\rm eff}=-1$ occurs when $\dot H=0$, that is,
$t=t_s/2$. Note that

\bea \label{any15} \frac{\ddot a}{a}
&=&\frac{16t_s}{27 \left(w_m + 1\right)^3\left(t_s - t\right)^2
t^2} \nn && \times \left\{ t - \frac{\left(3w_m +
1\right)t_s}{4}\right\}\ . \eea
Hence, if $w_m>-1/3$, the deceleration of the universe turns to the acceleration at $t= t_a \equiv \left(3w_m + 1\right)t_s/4$.

As second example, we now consider

\be \label{any30} f(t)=H(t)=\frac{\alpha\beta t} {\kappa^2 \left( 1 -
\frac{\beta}{2}\ln \left(\gamma +
\frac{t^2}{\kappa^2}\right)\right) \left(\gamma +
\frac{t^2}{\kappa^2}\right)}\ . \ee Since \bea \label{any31}
\frac{\ddot a}{a}&=&\frac{\alpha\beta} {\kappa^2 \left( 1 -
\frac{\beta}{2}\ln \left(\gamma +
\frac{t^2}{\kappa^2}\right)\right)^2 \left(\gamma +
\frac{t^2}{\kappa^2}\right)^2} \nn && \times \left\{ \left( 1 -
\frac{\beta}{2}\ln \left(\gamma +
\frac{t^2}{\kappa^2}\right)\right) \left(\gamma -
\frac{t^2}{\kappa^2}\right) \right. \nn && \left. +
\frac{\beta(1+\alpha)t^2}{\kappa^2}\right\}\ , \eea
if $t>0$, there are two solutions of $\ddot a=0$ , one corresponds to late time and another corresponds to early time. The late time solution $t=t_l$ of $\ddot a=0$ is obtained by neglecting $\gamma$. One obtains

\be
\label{any24} t=t_l \sim \kappa\e^{1/\beta - \alpha -1}\ .
\ee
On the other hand, the early time solution could be found by neglecting $\beta$, which is ${\cal O}\left(10^{-2}\right)$, to be

\be \label{SN5} t=t_e\sim \kappa \sqrt{\gamma}\ .
\ee
Then the universe undergoes accelerated expansion when $0<t<t_e$ and $t_l<t<t_s$. Here $t_s$ is Rip time:

\be
\label{BRt} t_s=\kappa \sqrt{-\gamma + \e^{2/\beta}}\sim \kappa \e^{1/\beta}\ .
\ee
$t_e$ may be identified with the time when the inflation ended. One is able to define the number of the e\,-\,foldings $N_e$ as

\be
\label{SN6} N_e = \ln
\left(a\left(t_e\right)/a\left(0\right)\right)\ .
\ee
Hence, it follows

\be \label{any36} N_e= -\alpha \ln \left(\frac{1-\frac{\beta}{2}\ln ( 2\gamma )}
{1-\frac{\beta}{2}\ln ( \gamma )}\right)\ .
\ee
Thus, we demonstrated that generalized EoS of the universe, in the same way as scalar model of Ref.\cite{any,uni}  may present the natural unification of the early time inflation and late time acceleration. Working in the same direction, with more observational constraints, one can suggest even more realistic generalized EoS of the universe, describing the cosmological evolution in great detail.

However, some comments are necessary at this point. As for the specific models analyzed above, to test models against observations is convenient to translate all in terms of $z$, i.e.

\be
H(z)=f(z,\dot{z})=-\frac{\dot{z}}{z+1}\,,
\ee
and then $E(z)$. This allows to select interesting ranges of $z$ to compare withobservations, for example $100< z < 1000$ for very far universe (essentially CMBR data), $10 < z < 100$ (structure formation), $0 < z < 10$ (present universe probed by standard candles, lookback time, etc.). Then different datasets, coming from different observational campaigns, have to be consistently matched with the same cosmological solution ranging from inflation to present accelerated era. This program  could be hard to be realized in details  because of the difficulties to join together a reasonable patchwork of data coming from different epochs, but it is possible in principle.

\section{Constraining the models}

Having in mind the theoretical considerations of last subsection, let us now develop a method to constrain dark energy models with generalized EoS against observations. In particular, we will use data coming from SNeIa and radio-galaxies for the above $IM$ and $QE$ models which reproduce the present day cosmic acceleration by means of inhomogeneous corrections in the dark energy EoS.

\subsection{The method}

In order to constrain the EoS characterizing parameters, we maximize the following likelihood function\,:

\begin{equation}
{\cal{L}} \propto \exp{\left [ - \frac{\chi^2({\bf p})}{2} \right ]}
\label{eq: deflike}
\end{equation}
where {\bf p} denotes the set of model parameters and the pseudo\,-\,$\chi^2$ merit function is defined as\,:

\begin{eqnarray}
\chi^2({\bf p}) & = & \sum_{i = 1}^{N}{\left [ \frac{y^{th}(z_i, {\bf p}) - y_i^{obs}}{\sigma_i} \right ]^2} \nonumber \\
~ & + & \displaystyle{\left [ \frac{{\cal{R}}({\bf p}) - 1.716}{0.062} \right ]^2} + \displaystyle{\left [ \frac{{\cal{A}}({\bf p}) - 0.469}{0.017} \right ]^2}  \ .
\label{eq: defchi}\
\end{eqnarray}
Let us discuss briefly the different terms entering Eq.(\ref{eq: defchi}). In the first one, we consider the dimensionless coordinate distance $y$ to an object at redshift $z$ defined as\,:

\begin{equation}
y(z) = \int_{0}^{z}{\frac{dz'}{E(z')}}
\label{eq: defy}
\end{equation}
and related to the usual luminosity distance $D_L$ as $D_L = (1 + z) r(z)$. Daly \& Djorgovki \cite{DD04} have compiled a sample comprising data on $y(z)$ for the 157 SNeIa in the Riess et al. \cite{Riess04} Gold dataset and 20 radio-galaxies from \cite{RGdata}, summarized in Tables\,1 and 2 of \cite{DD04}. As a preliminary step, they have fitted the linear Hubble law to a large set of low redshift ($z < 0.1$) SNeIa thus obtaining\,:

\begin{displaymath}
h = 0.664 {\pm} 0.008 \ .
\end{displaymath}
We thus set $h = 0.664$ in order to be consistent with their work, but we have checked that varying $h$ in the $68\%$ CL quoted above does not alter the main results. Furthermore, the value we are using is consistent also with $H_0 = 72 {\pm} 8 \ {\rm km \ s^{-1} \ Mpc^{-1}}$ given by the HST Key project \cite{Freedman} based on the local distance ladder and with the estimates coming from the time delay in multiply imaged quasars \cite{H0lens} and the Sunyaev\,-\,Zel'dovich effect in X\,-\,ray emitting clusters \cite{H0SZ}.

The second term in Eq.(\ref{eq: defchi}) makes it possible to extend the redshift range over which $y(z)$ is probed resorting to the distance to the last scattering surface. Actually, what can be determined from the CMBR anisotropy spectrum is the so called {\it shift parameter} defined as \cite{WM04,WT04}\,:

\begin{equation}
{\cal R} \equiv \sqrt{\Omega_M} y(z_{ls})
\label{eq: defshift}
\end{equation}
where $z_{ls}$ is the redshift of the last scattering surface which can be approximated as \cite{HS96},:

\begin{equation}
z_{ls} = 1048 \left ( 1 + 0.00124 \omega_b^{-0.738} \right ) \left ( 1 + g_1 \omega_M^{g_2} \right )
\label{eq: zls}
\end{equation}
with $\omega_i = \Omega_i h^2$ (with $i = b, M$ for baryons and total matter respectively) and $(g_1, g_2)$ given in Ref.\,\cite{HS96}. The parameter $\omega_b$ is well constrained by the baryogenesis calculations contrasted to the observed abundances of primordial elements. Using this method, Kirkman et al. \cite{Kirk} have determined\,:

\begin{displaymath}
\omega_b = 0.0214 {\pm} 0.0020 \ .
\end{displaymath}
Neglecting the small error, we thus set $\omega_b = 0.0214$ and use this value to determine $z_{ls}$. It is worth noting, however, that the exact value of $z_{ls}$ has a negligible impact on the results and setting $z_{ls} = 1100$ does not change  constraints on the other model parameters.

Finally, the third term in the definition of $\chi^2$ takes into account the recent measurements of the {\it acoustic peak} in the large scale correlation function at $100 \ h^{-1} \ {\rm Mpc}$ separation detected by Eisenstein et al. \cite{Eis05} using a sample of 46748 luminous red galaxies (LRG) selected from the SDSS Main Sample \cite{SDSSMain}. Actually, rather than the position of acoustic peak itself, a closely related quantity is better constrained from these data defined as \cite{Eis05}\,:

\begin{equation}
{\cal{A}} = \frac{\sqrt{\Omega_M}}{z_{LRG}} \left [ \frac{z_{LRG}}{E(z_{LRG})} y^2(z_{LRG}) \right ]^{1/3}
\label{eq: defapar}
\end{equation}
with $z_{LRG} = 0.35$ the effective redshift of the LRG sample. As it is clear, the ${\cal{A}}$ parameter depends not only on the dimensionless coordinate distance (and thus on the integrated expansion rate), but also on $\Omega_M$ and $E(z)$ explicitly which removes some of the degeneracies intrinsic in distance fitting methods. Therefore, it is particularly interesting to include ${\cal{A}}$ as a further constraint on the model parameters using its measured value \cite{Eis05}\,:

\begin{displaymath}
{\cal{A}} = 0.469 {\pm} 0.017 \ .
\end{displaymath}
Note that, although similar to the usual $\chi^2$ introduced in statistics, the reduced $\chi^2$ (i.e., the ratio between the $\chi^2$ and the number of degrees of freedom) is not forced to be 1 for the best fit model because of the presence of the priors on ${\cal{R}}$ and ${\cal{A}}$ and since the uncertainties $\sigma_i$ are not Gaussian distributed, but take care of both statistical errors and systematic uncertainties. With the definition (\ref{eq: deflike}) of the likelihood function, the best fit model parameters are those that maximize ${\cal{L}}({\bf p})$. However, to constrain a given parameter $p_i$, one resorts to the marginalized likelihood function defined as\,:

\begin{equation}
{\cal{L}}_{p_i}(p_i) \propto \int{dp_1 \ldots \int{dp_{i - 1} \int{dp_{i + 1} ... \int{dp_n {\cal{L}}({\bf p})}}}}
\label{eq: defmarglike}
\end{equation}
that is normalized at unity at maximum. Denoting with $\chi_0^2$ the value of the $\chi^2$ for the best fit model, the $1 \ {\rm and} \ 2 \sigma$ confidence regions are determined by imposing $\Delta \chi^2 = \chi^2 - \chi_0^2 = 1$ and $\Delta \chi^2 = 4$ respectively.

\begin{table}
\begin{center}
\begin{tabular}{|c|c|c|c|}
\hline
Par & $bf$ & $1 \sigma$ CR & $2 \sigma$ CR \\
\hline  \hline
$w_{DE}$ & $-1.03$ & $(-1.18, -0.91)$ & $(-1.39, -0.82)$ \\
$\tilde{\Omega}_M$ & $0.31$ & $(0.27, 0.36)$ & $(0.23, 0.41)$ \\
$\Omega_M$ & $0.29$ & $(0.27, 0.31)$ & $(0.25, 0.33)$ \\
\hline
$w_M^{eff}$ & $0.018$ & $(-0.022, 0.081)$ & $(-0.068, 0.142)$ \\
$w_X(z = 0)$ & $-1.13$ & $(-1.39, -0.92)$ & $(-1.71, -0.76)$ \\
$q_0$ & $-0.54$ & $(-0.71, -0.41)$ & $(-0.94, -0.31)$ \\
$z_T$ & $0.61$ & $(0.51, 0.71)$ & $(0.41, 0.83)$ \\
$t_0$ (Gyr) & $14.08$ & $(13.45, 14.71)$ & $(12.90, 15.44)$ \\
\hline
\end{tabular}
\end{center}
\caption{Summary of the results of the likelihood analysis for the $IM$ model. The maximum likelihood value ($bf$) and the $1$ and $2 \sigma$ confidence range (CR) for the model parameters $(w_{DE}, \tilde{\Omega}_M, \Omega_M)$ and some derived quantities are reported.}
\end{table}

\subsection{Results for the $IM$ model}

Before discussing the results of the likelihood analysis, it is worth stressing that the method described above makes it possible to obtain constraints on $\Omega_M$ directly and not only on $\tilde{\Omega}_M$. Actually, since we have used, in our analysis, only the dimensionless coordinate distance, the likelihood function ${\cal{L}}$ should depend only on the parameters entering $E(z)$ which determines the dimensionless coordinate distance through Eq.(\ref{eq: defy}). As a consequence, only $\tilde{\Omega}_M$ and $w_{DE}$ could be constrained without any possibility to infer $\Omega_M$. The use of the priors on the shift ${\cal{R}}$ and acoustic peak parameters ${\cal{A}}$ is the way to break the degeneracy between $\Omega_M$ and the ratio $w_{M}^{eff}/w_{DE}$ because both ${\cal{R}}$ and ${\cal{A}}$ depends explicitly on $\Omega_M$ and not implicitly through the (integrated or not) Hubble parameter. This example stresses the need for going beyond the SNeIa Hubble diagram in order to not only lessen the impact of eventual systematic errors, but also directly break possible theoretical degeneracies among model parameters. Table\,I reports a summary of the results giving best fit values at $1$ and $2 \sigma$ confidence ranges for some interesting derived quantities. Since the uncertainties on $(w_{DE}, \tilde{\Omega}_M, \Omega_M)$ are not Gaussian distributed, we do not apply a naive propagation of errors to determine the constraints on each derived quantity. We thus estimate the $1$ and $2 \sigma$ confidence ranges on the derived quantities by randomly generating 20000 points in the space of model parameters using the marginalized likelihood functions of each parameter and then deriving the likelihood function of the corresponding quantity. This procedure gives a conservative estimate of the uncertainties which is enough for our aims. Note that this procedure may also be used if an analytical expression is not available.

\begin{figure}
\centering \resizebox{8.5cm}{!}{\includegraphics{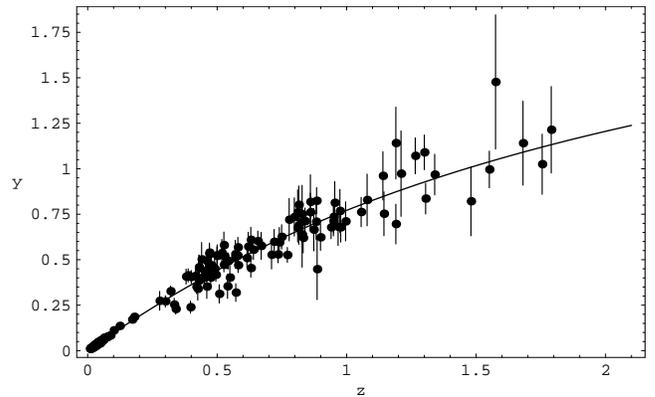}}
\caption{The best fit curve superimposed on the observed dimensionless coordinate distance $y(z)$ for $IM$ model.}
\label{fig: bfa}
\end{figure}

The best fit model parameters gives a theoretical $y(z)$ diagram which is in very good agreement with the observed data as convincingly shown in Fig.\,\ref{fig: bfa}. It is worth noting that the effective dark energy EoS $w_{DE}$ is quite close to the $\Lambda$CDM value which is not surprising since this latter model is known to fit very well a larger set of observations (including also CMBR anisotropy spectrum and matter power spectrum \cite{Teg03,Sel04}). For the same reason, it is also not unexpected that $\Omega_M$ turns out to be perfectly consistent with estimates coming from, e.g., WMAP CMBR spectrum \cite{WMAP} and clustering data \cite{LSS}. As a result, the best fit $\tilde{\Omega}_M$ is only slightly larger than $\Omega_M$ thus suggesting that $w_M^{eff}$ is vanishingly small. As reported in Table\,I, the best fit $w_M^{eff}$ is indeed very small and $w_M^{eff} = 0$ is well within the $1 \sigma$ confidence range. Based on Eq.(\ref{eq: defwde}), one may argue that $w_H$ is practically null and hence conclude that there is no need of any $H^2$ correction term in the dark energy EoS. However, such a result is biased by a theoretical prejudice. Actually, if we assume that Eq.(\ref{eq: wavsz}) is correct, then the present day EoS turns out to be in the phantom region $w_X < -1$ and, using Eq.(\ref{eq: estwf}), the true barotropic factor $w_f$ is estimated (at $2 \sigma$ CR) as\,:

\begin{displaymath}
-1.35 \le w_f \le -0.75
\end{displaymath}
which is still more in the domain of phantom fields. A possible way to recover $w_f > -1$ without altering none of the fit results is to resort to Eq.(\ref{eq: pabis}) rather than Eq.(\ref{eq: pa}). In such an approach, while the present day value of the dark energy EoS $w_X(z = 0)$ is still given by the value reported in Table\,I, the constraints above on $w_f$ are no more valid and should be replaced by\,:

\begin{table}
\begin{center}
\begin{tabular}{|c|c|c|c|}
\hline
Par & $bf$ & $1 \sigma$ CR & $2 \sigma$ CR \\
\hline  \hline $q_0$ & $-0.54$ & $(-0.66, -0.48)$ & $(-0.78,
-0.42)$ \\ $E_s$ & $0.941$ & $(0.937, 0.945)$ & $(0.924, 0.988)$
\\ $\Omega_M$ & $0.28$ & $(0.26, 0.29)$
& $(0.24, 0.31)$ \\
\hline $w_X(z = 0)$ & $-0.93$ & $(-1.0, -0.71)$ & $(-1.0, -0.50)$
\\ $z_T$ & $0.70$ & $(0.62, 0.76)$ & $(0.48, 0.82)$ \\ $t_0$ (Gyr) &
$14.21$ &
$(13.89, 14.56)$ & $(13.37, 14.85)$ \\
\hline
\end{tabular}
\end{center}
\caption{Summary of the results of the likelihood analysis for the $QE$ model. The maximum likelihood value ($bf$) and the $1$ and $2 \sigma$ confidence range (CR) for the model parameters $(q_0, E_s, \Omega_M)$ and some derived quantities are reported.}
\end{table}

 \begin{displaymath}
-1.35 \le w_f \left ( 1 - \frac{w_{dH} \kappa^2}{2} \right )^{-1} \le -0.75 \ .
\end{displaymath}
It is then possible to fit the same dataset with the same accuracy by choosing a whatever value of $w_f > -1$ provided that $w_{dH}$ is then set in such a way that the above constraint is not violated.

Having discussed the constraints on the dark energy EoS, it is interesting to compare some model predictions with the current estimates in literature. As a first test, we consider the present day value $q_0$ of the deceleration parameter. Since, as a general rule, $q_0 = 1/2 + (3/2) \Omega_X w_x(z = 0)$ and we get that both $\Omega_X = 1 - \Omega_M$ and $w_x(z = 0)$ are consistent with the $\Lambda$CDM values, it is not surprising that the best fit $q_0 = -0.54$ is quite similar to the concordance model prediction $q_0 \simeq -0.55$ \cite{WMAP,Riess04,Teg03,Sel04}. On the other hand, $q_0$ may also be determined in a model independent way by using a cosmographic approach, i.e. expanding the scale factor $a(t)$ in a Taylor series and fitting to distance related data. Using a fifth order expansion and the Gold SNeIa sample, John \cite{moncy} has obtained $q_0 = -0.90 {\pm} 0.65$ which is a so large range to be virtually in agreement with almost everything. However, we note that our best fit value is significantly larger than his estimate. Although this could be a problem, we do not consider it particularly worrisome given the uncertainties related to the impact of truncating the expansion of $a(t)$ to a finite order.

According to the most recent estimates, the cosmic speed up has started quite recently so that it is possible to detect the first signature of the past decelerated expansion. This opens the way to determine the transition redshift $z_T$ defined as the solution of the equation $q(z_T) = 0$. For the $IM$ model, it is possible to derive an analytical expression\,:

\begin{equation}
z_T = \left [ \frac{1 - 2 q_0 + 3 w_{DE}}{(1 - 2 q_0) (1 + w_{DE})} \right ]^{\frac{1}{3 w_{DE}}} - 1 \ .
\label{eq: zta}
\end{equation}
The constraints we get are summarized in Table\,I. As a general remark, we note that the best fit value $z_T = 0.61$ is once again close to that for the concordance $\Lambda$CDM model for the same reasons explained above for $q_0$. Most of the $z_T$ estimates available in literature are model dependent since they have been obtained as a byproduct of fitting an assumed model (typically the $\Lambda$CDM one) to the data. A remarkable exception is the reported $z_T = 0.46 {\pm} 0.13$ found by Riess et al. \cite{Riess04} fitting the ansatz $q(z) = q_0 + (dq/dz)_{z = 0} z$ to the Gold SNeIa sample. Although our best fit estimate is formally excluded at $1 \sigma$ level by the result of Riess et al., there is a wide overlap between our estimated $1 \sigma$ CR and that of Riess et al. so that we may safely conclude that the two results agree each other.

\begin{figure}
\centering \resizebox{8.5cm}{!}{\includegraphics{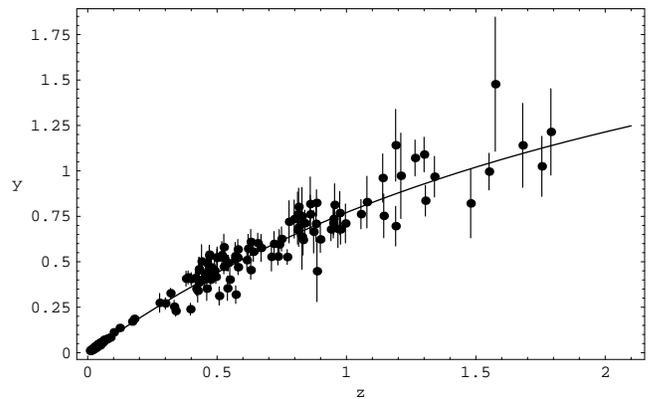}}
\caption{The best fit curve superimposed on the observed dimensionless coordinate distance $y(z)$ for the $QE$ model.}
\label{fig: bfb}
\end{figure}

As a final check, we consider the present age of the universe that may be evaluated as\,:

\begin{equation}
t_0 = t_H \int_{0}^{\infty}{\frac{dz}{(1 + z) E(z)}}
\label{eq: deftz}
\end{equation}
with $t_H = 1/H_0 = 9.778 h^{-1} \ {\rm Gyr}$ the Hubble time. Our best fit estimate turns out to be $t_0 = 14.08 \ {\rm Gyr}$ with the $2 \sigma$ CR extending from $12.90$ up to $15.44 \ {\rm Gyr}$. This result is in satisfactory agreement with previous model dependent estimates such as $t_0 = 13.24_{+0.41}^{+0.89} \ {\rm Gyr}$ from Tegmark et al. \cite{Teg03} and $t_0 = 13.6 {\pm} 0.19 \ {\rm Gyr}$ given by Seljak et al. \cite{Sel04}. Aging of globular clusters \cite{Krauss} and nucleochronology \cite{Cayrel} give model independent (but affected by larger errors) estimates of $t_0$ still in good agreement with our one. Note also that, the larger value of the best fit $t_0$ we obtain is also a consequence of our assumption $h = 0.664$ that is smaller than $h = 0.71$ used by Tegmark et al. and Seljak et al. as a result of their more elaborated and comprehensive data analysis.

\subsection{Results for $QE$ model}

The results of the likelihood analysis obtained in this case are summarized in Table\,II where constraints on the model parameters $(q_0, E_s, \Omega_M)$ and on some derived quantities are given, while Fig.\,\ref{fig: bfb} shows the best fit curve superimposed to the data. As it is clear, the agreement is quite good and, to a large extent, the $QE$ model works as well as the $IM$ one in fitting the used dataset.

\begin{figure}
\centering \resizebox{8.5cm}{!}{\includegraphics{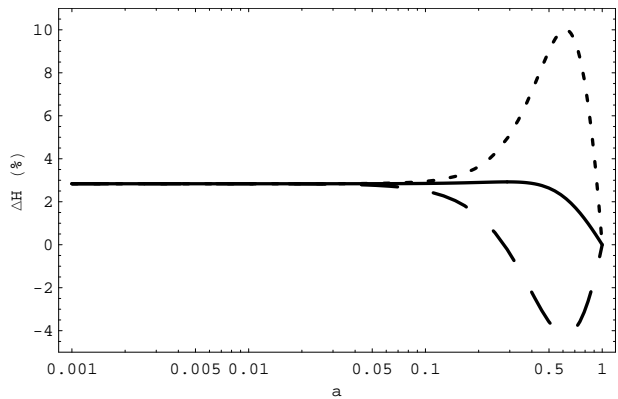}}
\caption{The percentage difference $\Delta H$ with respect to the $\Lambda$CDM scenario for the $IM$ model. We set the parameters $\Omega_M$ (for both the $IM$ and the $\Lambda$CDM model) and $\tilde{\Omega}_M$ to their best fit values in Table\,I, and choose three values for $w_{DE}$, namely $w_{DE} = -1.39$ (short dashed), $-1.03$ (solid) and $-0.82$ (long dashed).}
\label{fig: imvslcdm}
\end{figure}

Although the dark energy EoS is significantly different, it is interesting to note that most of the $QE$ model predicted quantities are in very good agreement with the same quantities for the $IM$ scenario. It is worthwhile to spend some words on the present day value of the deceleration parameter. While for the $IM$ model $q_0$ is a derived quantity, here $q_0$ is used to assign the model characteristics and is therefore obtained directly from the likelihood procedure. Notwithstanding the different role played in the two cases, the best fit value is identical, while the $1$ and $2 \sigma$ CR overlap very well\footnote{It is not surprising that the $1$ and $2 \sigma$ CR on $q_0$ in Table\,I are slightly larger than those in Table\,II. Actually, for the $IM$ model, the constraints on $q_0$ are determined by the uncertainties on the other model parameters, while, in the present case, they are obtained from the fitting procedure directly.}. This result may be qualitatively explained noting that, in both cases, setting the model parameters to their best fit values leads to a cosmological model that is quite close (from a dynamical point of view) to the concordance $\Lambda$CDM scenario. As a further support to this picture, let us note that $w_X(z = 0) \simeq -1$ as for the cosmological constant. Note that, because of our choice to be in a non-phantom regime, $w_X$ is forced to be larger than $-1$ so that we may give only upper limits on this quantity. If we had chosen  the phantom regime, and the possibility to cross the phantom divide, we should have given also lower limits to $w_X$.

The fact that the model is indeed close to the usual $\Lambda$CDM can be also understood looking at the constraints on $E_s$. Eq.(\ref{eq: wb}) shows that the $\Lambda$CDM model is obtained in the limit $E_s = 1$ so that, in a certain sense, this parameter measures how far the dark energy fluid is from a simple cosmological constant. Indeed, the best fit value $E_s = 0.941$ is quite close to 1 so that one could argue that only minor deviations from the standard $\Lambda$CDM model are detected.

As a final remark, we note that both the transition redshift $z_T$ (that, in this case, has to be evaluated numerically) and the age of the universe $t_0$ are in agreement with what is expected for the $\Lambda$CDM model (once the value of $h$ is raised to the commonly adopted $h = 0.70$) so that it is not surprising that we find good agreement with the values reported in literature and quoted above when discussing the results for the $IM$ model.

\begin{figure}
\centering \resizebox{8.5cm}{!}{\includegraphics{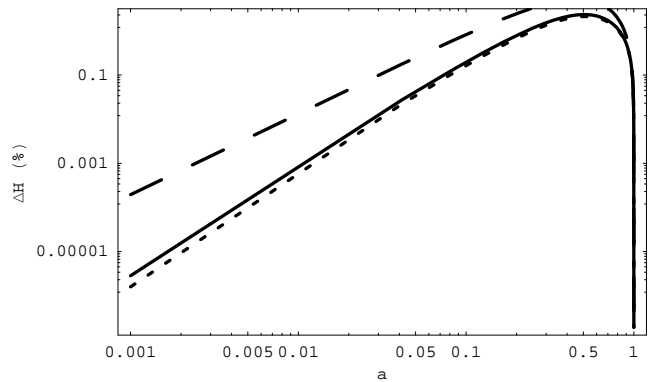}}
\caption{The percentage difference $\Delta H$ with respect to the $\Lambda$CDM scenario for the $QE$ model. We set the parameters $\Omega_M$ (for both the $QE$ and the $\Lambda$CDM model) and $q_0$ to their best fit values in Table\,II, and choose three values for $E_s$, namely $E_s = 0.924$ (short dashed), $0.941$ (solid) and $0.988$ (long dashed).}
\label{fig: qevslcdm}
\end{figure}

\section{High redshift behaviour}

The likelihood analysis presented above has demonstrated that both the $IM$ and the $QE$ model are able to fit the data on the dimensionless coordinate distance to SNeIA and radio\,-\,galaxies. As a further test, we have also imposed priors on the shift and acoustic peak parameters in order to better constrain the models. Nevertheless, the bulk of the data we have considered only probes the low redshift ($z \le 1.7$) epoch of the universe expansion. It is therefore worth investigating what are the consequences of the bulk viscosity corrections to the dark energy EoS at higher redshifts. We can infer some interesting qualitative results by simply comparing the high redshift expansion rate to those predicted by the successful concordance $\Lambda$CDM model.

To this aim, we show in Figs.\,\ref{fig: imvslcdm} and \ref{fig: qevslcdm} the percentage deviation $\Delta H = |(H_{\Lambda} - H_{mod})/H_{\Lambda}|$ of the Hubble rate $H(a)$ predicted by the $IM$ and $QE$ models respectively with respect to the $\Lambda$CDM scenario. As it is apparent, in both cases, $\Delta H$ significantly differs from the null value only in the very recent epochs, i.e. on the redshift range probed by the SNeIA data.

Actually, these results may be qualitatively explained for both models. As Eq.(\ref{eq: ea}) shows, the Hubble parameter for the $IM$ model is formally the same as that of the $\Lambda$CDM model provided the matter content $\Omega_M$ is shifted upwards being replaced by $\tilde{\Omega}_M$. According to the constraints discussed above, however, $\Omega_M$ and $\tilde{\Omega}_M$ are almost the same and, furthermore, $w_{DE}$ is only slightly different from the cosmological constant value. It is thus not surprising at all that the $IM$ and the $\Lambda$CDM models predict almost the same expansion history. A similar discussion also holds for the $QE$ case. As yet said, the $QE$ model reduces to the $\Lambda$CDM one in the limit $E_s \rightarrow 1$ and indeed, the $1 \sigma$ CR for this parameter is not too far from 1 to give rise to significant deviations at high redshift.

Although encouraging, both these results are not surprising. Indeed, any reasonable dark energy model should predict that the energy density of this component becomes vanishingly small in the far past in order to allow the universe undergoing a matter dominated phase. Moreover, the dark energy contribute to the total energy budget must be negligible during the radiation dominated period so that the nucleosynthesis can take place in the usual way. As such, Figs.\,\ref{fig: imvslcdm} and \ref{fig: qevslcdm} are reassuring since they show that both the $IM$ and $QE$ models behave as the $\Lambda$CDM one during the matter and radiation dominated epoch thus making us confident that the constraints from nucleosynthesis are satisfied.

\section{Structure formation}

Including viscosity terms in the dark energy EoS could impact also
the formation of cosmological structures. It is common to
investigate the evolution of matter density perturbations $\delta
= \delta \rho_M/\rho_M$, in the linear regime, solving\,:

\begin{equation}
\ddot{\delta} + 2 H \dot{\delta} - 4 \pi G \rho_M \delta = 0 \ .
\label{eq: perttime}
\end{equation}
This relation assumes that the dark energy does not cluster on
small scales so that $\rho_X$ concurs to determining the evolution
of $\delta$ only through the Hubble parameter. Moreover, the
derivation of Eq.(\ref{eq: perttime}) implicitly assumes that the
General Relativity in its Einsteinian formulation is the correct
theory of gravity (see, e.g., \cite{padbook} for a textbook
description).

The analysis of our generalized EoS presented up to now mainly
rely on the dynamics of the cosmic expansion which can be
investigated also assuming that both the $IM$ and $QE$ models are
phenomenological descriptions of an unspecified theory. Put in
other words, we do not need to know what is the Lagrangian giving
rise to Eqs.(\ref{eq: fried1}) and (\ref{eq: fried2}) with the
dark energy EoS given by Eqs.(\ref{eq: padot}) and (\ref{eq: pb}).
In order to investigate the issue of structure formation, it is,
on the contrary, mandatory to have a Lagrangian description of
both models.

Let us first consider the $IM$ model in which case the EoS (in its
more general form) is given by Eq.(\ref{eq: padot}). In Appendix
B, we show that such a model could be derived by the following
action\,:

\begin{displaymath}
S = \int{d^4x \sqrt{-g} \left \{ \frac{R}{2 \kappa^2} -
\frac{\omega(\phi)}{2} \partial_\mu \partial^{\mu} \phi - V(\phi)
\right \}} \ .
\end{displaymath}
Such an action is similar to that of most of scalar field
quintessence models with $V(\phi)$ the self\,-\,interaction
potential, but now the sign of the kinetic energy is not fixed
from the beginning. It is possible to show that this choice allows
to cross the phantom divide line in a natural way which is not the
case in the usual quintessence models. For what we are concerned
here, it is worth stressing that such an action makes it possible
to interpret the $IM$ model in the framework of Einsteinian
General Relativity\footnote{For a different interpretation, see
\cite{mengnew} where it is shown that Eq.(\ref{eq: padot}) may be
derived adding a bulk viscosity term in the energy\,-\,momentum
tensor.}. In this case, a leading role in determining the
clustering properties of the dark energy is played by the the EoS
$w_x$ and sound speed $c_s^2$. Assuming the pressure perturbations
are adiabatic, the sound speed reads\,:

\begin{equation}
c_s^2 = \frac{\partial p_X}{\partial \rho_X} = w_X + \left(
\frac{1}{\eta} \frac{d\eta}{dz} \right )^{-1} \frac{dw_X}{dz}
\label{eq: csdef}
\end{equation}
with $\eta = \rho_X(z)/\rho_{crit}$. For the $IM$ model, after
some algebra, we get\,:

\begin{equation}
c_s^2 = \frac{[(1 + z)^{-3} \eta - (\tilde{\Omega}_M - \Omega_M )] (1 + w_{DE}) w_{DE}}
{(1 + w_{DE}) (1 + z)^{-3} \eta - (\tilde{\Omega}_M - \Omega_M) w_{DE}}
\label{eq: csim}
\end{equation}
with\,:

\begin{equation}
\eta = (\tilde{\Omega}_M - \Omega_M)(1 + z)^3 + (1 -
\tilde{\Omega}_M)(1 + z)^{3(1 + w_{DE})} \ . \label{eq: etaim}
\end{equation}
It is immediately clear that, in the limit $w_M^{eff} = 0$, we get
$\tilde{\Omega}_M = \Omega_M$ and both the EoS and the sound speed
reduces to that of a quiessence model with constant EoS, i.e. we
get $c_s^2 = w_X = w_{DE}$. Indeed, for the best fit parameters,
$c_s^2$ departs from the constant value $w_DE$ only marginally
since $w_M^{eff}$ is quite small. Moreover, if we consider the
range $w_{DE} \le -1$, which is compatible with the constraints in
Table\,I, one may easily check that, for $z >> 0$, it is\,:

\begin{displaymath}
\eta(z) \simeq (1 - \tilde{\Omega}_M) (1 + z)^{3(1 + w_{DE}} \ \ , \ \ c_s^2 \simeq w_{DE} \ ,
\end{displaymath}
i.e. the $IM$ model reduces to a phantom one with constant EoS
$w_X = w_{DE}$. It is easy to show that the same approximate
result for the sound speed also holds in the case $w_{DE} \ge -1$.
These analogies make us confident that, in the $IM$ scenario, the
evolution of matter density perturbations could be studied as in
standard quiessence models (see, e.g., \cite{perc} and references
therein), but discussing further this issue is outside our aims.

Let us now consider the $QE$ model which, as shown in Appendix B,
could be derived by the same action above choosing a different
expression for $\omega(\phi)$. As a consequence, we can still use
the standard theory of perturbation. Starting from Eq.(\ref{eq:
pb}), we get\,:

\begin{equation}
c_s^2 = w_X + \frac{C_s E_s E' [E^2 - \Omega_M (1 + z)^3]}{2(1 +
w_X) E^2 [2 E E' - 3 \Omega_M (1 + z)^2]} \label{eq: csqe}
\end{equation}
with $w_X$ and $C_s$ given by Eqs.(\ref{eq: wb}) and (\ref{eq:
csqz}) respectively, $E = H(z)/H_0$ and the prime denoting
derivative with respect to $z$. Integrating Eq.(\ref{eq: deb}) for
a given values of the model parameters inserting the result into
Eq.(\ref{eq: csqe}) above, we find out that $c_s^2(z)$ has an
unpleasant behaviour since it starts from values larger than $1$,
which is not physically reasonable\footnote{Remember that we are
working in units with $c = 1$ so that Eq.(\ref{eq: csdef})
actually refers to $c_s^2/c^2$.}, and then takes negative values
for $z > z_c$, with $z_c$ in the range $(0.5, 1.5)$ depending on
$(q_0, E_s, \Omega_M)$. Positive values of the sound speed may be
obtained also for the generalized Chaplygin gas and give rise to
exponential blow up of the perturbations in the linear regime
\cite{sandvik}. However, it has been suggested that considering
entropy perturbations could solve the problem \cite{reis}. Even if
the $QE$ model significantly differs from the generalized
Chaplygin gas, it is possible that a similar mechanism could alter
the sound speed in such a way to eliminate the disturbing feature
of superluminal values. Indeed, when entropy perturbations enter
the game, pressure perturbations are no more adiabatic so that
Eq.(\ref{eq: csdef}) does not hold anymore. Whether this can also
regularize the sound speed for the $QE$ model is a topic worth to
be investigated, but it is outside our aims.

\section{Conclusions}

Dark energy EoS with inhomogeneous terms coming from geometry (e.g. $H, \dot{H}$) can yield cosmological models capable of avoiding shortcomings coming from coincidence problem and  a fine tuned sudden passage of the universe from the decelerated regime (dark matter dominated) to the today observed accelerated regime (dark energy dominated). Furthermore, such models allow to recover also early accelerated regimes with the meaning of inflationary behaviors. In this paper, we have constrained two physically relevant models using SNeIa and radio-galaxies data including priors on the shift and the acoustic peak parameters.

In the first class of models, referred to as $IM$ model, the dark energy pressure depends linearly not only on the  energy density, but also on $H^2$ and $\dot{H}$. Analytical expressions for $\rho_X(z)$, $w_X(z)$ and the dimensionless Hubble parameter $E(z)$ have been obtained. As an interesting result, we obtain for $E(z)$ the same functional expression as for the QCDM model provided the matter density parameter $\Omega_M$ and the dark energy EoS $w_X$ are replaced by biased quantities $\tilde{\Omega}_M$ and $w_{DE}$. As a consequence, fitting to observables only depending on $E(z)$ may lead to wrong results. In the second case, dubbed $QE$ model, the EoS is not given explicitly, but as the solution of a quadratic equation.  In particular, we have fitted the non-phantom regime $w_X \ge -1$.

The likelihood analysis presented in Sect.\,III has convincingly shown that both models are able to fit well the data on the dimensionless coordinate distance to SNeIa and radio\,-\,galaxies also taking into account the priors on the shift and acoustic peak parameters ${\cal{R}}$ and ${\cal{A}}$. Moreover, they predict values of the deceleration parameter $q_0$, the transition redshift $z_T$ and the age of the universe $t_0$ that are in agreement with most of the estimates available in literature. It is worth noting that, setting the model parameters to their best fit values, both the $IM$ and $QE$ models reproduce the same dynamics as the concordance $\Lambda$CDM scenario. Nevertheless, the underlying physical mechanism is radically different so that both approaches worth be further investigated.

On the one hand, we have here only considered the general {\it dynamics} of the universe, i.e. the Hubble parameter $H(z)$, which enters in the determination of distance related quantities. It is therefore mandatory to enlarge our attention considering other observational tests which probes the EoS itself rather than its integrated value. Ideal candidates are the growth index and the evolution of structures since $w_X(z)$ enters directly into the corresponding equations. Moreover, the full CMBR anisotropy spectrum has to be evaluated which is likely to offer the possibility to determine separately $(w_f, w_H, w_{dH})$ in the case of the $IM$ model. Actually, since our EoS are inhomogeneous, viscosity\,-\,like terms will appear in the equations governing the evolution of density perturbations thus probably leading to some distinctive signatures that could be detected.

On the other hand, further developments on the theoretical side are possible. Actually, Eqs.(\ref{eq: wabis}) and (\ref{eq: pb}) have been assumed rather than obtained from a fundamental theory. As shown in Appendix B and more deeply discussed in \cite{NO05}, it is possible to work out a modified gravity theory that leads to an effective dark energy fluid with EoS given by Eq.(\ref{eq: wabis}) or by Eq.(\ref{eq: pb}). It should therefore be interesting to investigate what constraints may be imposed on deviations from Einstein general relativity by the likelihood analysis presented here.

As it was demonstrated in \cite{NO05}, including more general viscosity term in the dark energy EoS brings new possibilities to construct both the early and late time universe. In particular, one may look for models that are able to give rise to both an early inflationary epoch and a present day accelerated expansion \cite{any}. Testing such models against observational data is the key to select among the different possibilities which is the most realistic one.

Finally some comments are necessary on the general philosophy of the classes of models studied here and on the constraints coming from the observations capable of restricting and selecting viable cosmologies. Considering also the general discussion in Sect.II (Subsect. C), it is clear that generalized EoS where further terms are added to $p=-\rho$ are preferable for the following reasons. A huge amount of observational evidences indicates that $\Lambda$CDM $(p=-\rho)$ is the cosmological scenario able to realistically describe better the today universe. Any evolutionary model passing from deceleration (dark matter dominance) to acceleration (dark energy dominance) should consistently reproduce, based on observations, such a scenario. In the $IM$ case which we have studied, adding  terms in Hubble parameter and its derivative allows always a comparison with standard matter parameters ($w_M$ and $\Omega_M$) which are, in some sense, "directly" observable. Then the number of arbitrary choices (for example, to fix priors) is not so large. On the other hand, $QE$ models, or in general implicit EoS, need several arbitrary choices which could result completely inconsistent to further and more refined observations. For example, in our case, we have arbitrarily discarded phantom-like regime which could result consistent with observations at large $z$ (far distances and early universe) and imposed arbitrary constraints on $q_0$. Due to these reasons, from an observational point of view, it is preferable to study models which imply corrections the $\Lambda$ EoS rather than giving EoS in implicit form.

\appendix

\section{Bulk viscosity}

Let $U^\mu$ be the four velocity of the cosmic fluid. Then the scalar expansion is given by $\theta\equiv U^\mu_{\ ;\mu}$ and we have $\theta=3H$.  When there is a bulk viscosity $\zeta$, which could be a function of the energy density $\rho_X$  of the fluid: $\zeta=\zeta\left(\rho_X\right)$, the first FRW equation is not changed $(3/\kappa^2)H^2=\rho$ but the second FRW equation and the conservation of the energy are  expressed as

\bea
\label{A1} 0&=&\frac{1}{\kappa^2}\left(2\dot H + 3H^2\right) + p_X - \zeta
\theta \nn &=&\frac{1}{\kappa^2}\left(2\dot H + 3H^2\right) + p_X
- 3H\zeta\left(\rho_X\right)\ , \\ \label{A2} 0&=&\dot \rho_X +
\theta \left(\rho_X + p_X - \zeta \theta\right) \nn &=&\dot \rho_X
+ 3H \left(\rho_X + p_X - 3H\zeta\left(\rho_X\right)\right)\ .
\eea
Then we obtain the effective pressure $\tilde p$ as

\be
\label{A3} \tilde p_X\equiv p_X - 3H\zeta\left(\rho_X\right)\ .
\ee
Then the second FRW equation and the conservation of the energy have the standard forms: expressed as

\bea \label{A4} 0&=&\frac{1}{\kappa^2}\left(2\dot H + 3H^2\right) + \tilde p_X \ , \\
\label{A5} 0&=&\dot \rho_X + 3H \left(\rho_X + \tilde p_X \right)
\ .
\eea
Then if $\rho_X$ and $p_X$ satisfy the EOS

\be \label{A6} F\left(\rho_X,p_X\right)=0\ , \ee
we obtain effectively generalized EOS:

\be
\label{A7} \tilde F\left(\rho_X,\tilde p_X,
H\right) \equiv F\left(\rho_X,\tilde p_X +
3H\zeta\left(\rho_X\right) \right)=0\ . \ee

\section{Theoretical foundation}

Here, we give two possible theoretical foundations for the $IM$
and $QE$ models discussed in the text. We may start with the
following action: \be \label{k1} S=\int d^4 x
\sqrt{-g}\left\{\frac{1}{2\kappa^2}R
    - \frac{1}{2}\omega(\phi)\partial_\mu \phi\partial^\mu \phi -
V(\phi)\right\}\ . \ee Then the energy density $\rho$ and the
pressure $p$ for the scalar field $\phi$ are given by \be
\label{k4} \rho = \frac{1}{2}\omega(\phi){\dot \phi}^2 + V(\phi)\
, \quad p = \frac{1}{2}\omega(\phi){\dot \phi}^2 - V(\phi)\ . \ee
Since we can always redefine the scalar field $\phi$ as $\phi\to
F(\phi)$ by an arbitrary function $F(\phi)$,  we can choose the
scalar field to be a time coordinate; $\phi=t$. Furthermore we
consider the case that $\omega(\phi)$ and $V(\phi)$ are given by a
single function $f(\phi)$ as \bea \label{any5} \omega(\phi) &=&-
\frac{2}{\kappa^2}f'(\phi) \ ,\nn V(\phi) &=&
\frac{1}{\kappa^2}\left(3f(\phi)^2 + f'(\phi)\right) \ . \eea Then
\bea \label{kk1} \rho&=& \frac{3}{\kappa^2}f(\phi)^2\ ,\nn p&=&
-\frac{3}{\kappa^2}f(\phi)^2 - \frac{2}{\kappa^2}f'(\phi)\ . \eea
Since $\rho=f^{-1}\left(\kappa\sqrt{\rho/3}\right)$, we reobtain
the generalized EoS \be \label{SN1bis} p=-\rho -
\frac{2}{\kappa^2}f'\left(f^{-1}\left(\kappa\sqrt{\frac{\rho}{3}}\right)\right)\
.  \ee As an example,  the case may be considered \be
\label{SN2bis} f=f_0\phi^2 + f_1\ , \ee with constants $f_0$ and
$f_1$. Then (\ref{SN1bis}) gives \be \label{SN3bis} 0=\left(p +
\rho\right)^2 - \frac{4f_0}{\kappa^4} f\left(1 -
\frac{f_1}{f}\right)\ . \ee If we neglect the contribution from
the matter, from the first FRW equation, we find \be \label{kk2}
f^2=\frac{\kappa^2}{3}\rho =H^2\ . \ee One may rewrite
(\ref{SN3bis}) as \be \label{kk3} 0=\left(p + \rho\right)^2 -
\frac{4f_0}{\kappa^3} \sqrt{\frac{\rho}{3}} \left(1 -
\frac{f_1}{H}\right)\ , \ee which may have a structure similar to
Eq.(\ref{eq: pb}). One may also consider the case that \be
\label{kkk1} f=\frac{f_0}{\phi}\ , \ee which gives \be
\label{kkk2} p=\left(-1 + \frac{2}{3f_0}\right)\rho\ . \ee Hence,
we obtain homogeneous equation of state with $w=-1 + {2}/{3f_0}$.
In case without matter,  by using the first and second FRW
equations, we can rewrite (\ref{kkk2}) as \be \label{kkk3}
p=\left(\frac{\beta}{3f_0\alpha} - 1\right)\rho +
\frac{2(1-\beta)}{f_0\alpha \kappa^2}H^2 +
\frac{2(1-\alpha)}{\kappa^2\alpha}\dot H\ . \ee Here $\alpha$ and
$\beta$ are constants. Then by identifying \be \label{kkk4} w_X=
\frac{\beta}{3f_0\alpha} - 1\ ,\quad w_H=
\frac{2(1-\beta)}{f_0\alpha \kappa^2}\ , \quad
w_{dH}=\frac{2(1-\alpha)}{\kappa^2\alpha}\ , \ee one gets
Eq.(\ref{eq: wabis}), while we get Eq.(\ref{eq: wa}) for
$\alpha=1$.

We may include matter with constant EoS parameter $w_m\equiv p_m /
\rho_m$ to the action (\ref{k1}). Here $\rho_m$ and $p_m$ are
energy density and pressure of the matter.  When the matter is
included, the FRW equations give \bea \label{Km1} && \omega(\phi)
{\dot \phi}^2 = - \frac{2}{\kappa^2}\dot H - \left(\rho_m +
p_m\right)\ ,\nn && V(\phi)=\frac{1}{\kappa^2}\left(3H^2 + \dot
H\right) - \frac{\rho_m - p_m}{2} \ . \eea By using the energy
conservation $\dot\rho_m + 3H\left(\rho_m + p_m\right)=0$, we find
$\rho_m=\rho_{m0} a^{-3(1+w_m)}$ with a constant $\rho_{m0}$. Then
if \bea \label{Km2} \omega(\phi) &=&- \frac{2}{\kappa^2}g''(\phi)
- \left(w_m + 1\right)g_0 \e^{-3(1+w_m)g(\phi)}\ ,\nn V(\phi) &=&
\frac{1}{\kappa^2}\left(3g'(\phi)^2 + g''(\phi)\right) \nn &&
+\frac{w_m -1}{2}g_0 \e^{-3(1+w_m)g(\phi)} \ , \eea we find a
solution \bea \label{Km3} && H=g'(t)\ \nn && \left(a=a_0
\e^{g(t)}\ ,\quad a_0\equiv
\left(\frac{\rho_{m0}}{g_0}\right)^{\frac{1}{3(1+w_m)}}\right) \ .
\eea Then we find \bea \label{Km4} \rho
&=&\frac{3}{\kappa^2}g'(\phi)^2 - g_0\e^{-3(1+w_m)g(\phi)}\ ,\nn p
&=&-\frac{1}{\kappa^2}\left(3g'(\phi)^2 + 2g''(\phi)\right) \nn &&
- w_m g_0\e^{-3(1+w_m)g(\phi)}\ . \eea Then by deleting $\phi$
from the above equations (\ref{Km4}), we obtain an EoS
corresponding to (\ref{SN1bis}).

We may also consider the possibility to obtain the generalized EoS
from the modified  gravity. As an illustrative example, the
following action is considered: \be \label{HD1} S=\int d^4 x
\sqrt{-g}\left(\frac{1}{2\kappa^2}R + {\cal L}_{\rm matter} +
f(R)\right)\ . \ee Here $f(R)$ can be an arbitrary function of the
scalar curvature $R$ and ${\cal L}_{\rm matter}$ is the Lagrangian
for the matter. In the FRW universe, the gravitational equations
are: \bea \label{HD2} 0&=& - \frac{3}{\kappa^2}H^2 + \rho -
f\left(R=6\dot H + 12 H^2\right) \nn && + 6\left(\dot H + H^2 - H
\frac{d}{dt}\right) \nn
&& \times f'\left(R=6\dot H + 12 H^2\right) \ ,\\
\label{HD3} 0&=& \frac{1}{\kappa^2}\left(2\dot H + 3H^2\right) + p
+ f\left(R=6\dot H + 12 H^2\right) \nn && + 2\left( - \dot H -
3H^2 + \frac{d^2}{dt^2} + 2H \frac{d}{dt}\right) \nn && \times
f'\left(R=6\dot H + 12 H^2\right)\ . \eea Here $\rho$ and $p$ are
the energy density and the pressure coming from ${\cal L}_{\rm
matter}$. They may satisfy the equation of state like $p=w\rho$.
One may now define the effective energy density $\tilde \rho$ and
$\tilde p$ by \bea \label{HD4} \tilde\rho &\equiv& \rho -
f\left(R=6\dot H + 12 H^2\right) \nn && + 6\left(\dot H + H^2 - H
\frac{d}{dt}\right) \nn
&& \times f'\left(R=6\dot H + 12 H^2\right) \ ,\\
\label{HD5} \tilde p&=& p + f\left(R=6\dot H + 12 H^2\right) \nn
&& + 2\left( - \dot H - 3H^2 + \frac{d^2}{dt^2} + 2H
\frac{d}{dt}\right) \nn && \times f'\left(R=6\dot H + 12
H^2\right)\ . \eea Thus, it follows \bea \label{HD6} \tilde p&=&
w\tilde \rho  + (1+w)f\left(R=6\dot H + 12 H^2\right) \nn && +
2\left( \left(-1 - 3w\right) \dot H - 3\left(1+w\right) H^2 +
\frac{d^2}{dt^2} \right.\nn && \left. + \left(2 + 3w\right) H
\frac{d}{dt}\right) f'\left(R=6\dot H + 12 H^2\right) \ . \eea
Especially if we consider the case that $f=f_0R$ with a constant
$f_0$, we obtain \be \label{kk4} \tilde p= w {\tilde \rho} +
6f_0\dot H + 6(1+w)f_0 H^2\ , \ee which reproduces \ref{eq:
wabis}) by identifying $\tilde \rho=\rho_X$, $\tilde p=p_X$,
$w=w_X$, $6f_0=w_{dH}$ $6(1+w)=w_H$. Furthermore if we can neglect
$\dot H$ compared with $H^2$, we also obtain Eq.(\ref{eq: wa}).

It is thus demonstrated that the generalized EoS considered in
this paper maybe derived also from scalar\,-\,tensor or fourth
order gravity theories.

\end{document}